\documentclass[aps,pra,twocolumn,superscriptaddress,amsmath,amssymb,showpacs]{revtex4-1}
\pdfoutput=1
\usepackage[caption=false]{subfig}
\usepackage{algorithm}
\usepackage{color}
\usepackage{algorithmic}
\usepackage{amsmath}
\usepackage{amssymb}
\usepackage{amsthm}
\usepackage{bm}
\usepackage{thmtools}

\usepackage{array}
\usepackage{url}
\usepackage{hyperref}
\usepackage{graphicx}
\usepackage{bibentry}
\usepackage{microtype}

\providecommand{\U}[1]{\protect\rule{.1in}{.1in}}
\newtheorem*{theorem*}{Theorem}

\newcommand{\ket}[1]{\left| #1 \right>} 

\begin{document}
\title{Efficient representation of Gaussian states for multi-mode non-Gaussian quantum state engineering via subtraction of arbitrary number of photons}
\author{Christos N. Gagatsos}
\affiliation{College of Optical Sciences, University of Arizona, 1630 E. University Blvd., Tucson, Arizona 85719, United States of America}
\author{Saikat Guha}
\affiliation{College of Optical Sciences, University of Arizona, 1630 E. University Blvd., Tucson, Arizona 85719, United States of America}
\affiliation{Department of Electrical and Computer Engineering, University of Arizona, 1230 E Speedway Blvd., Tucson, Arizona 85719, United States of America}

\begin{abstract}
We introduce a complete description of a multi-mode bosonic quantum state in the coherent-state basis (which in this work is denoted as ``$K$" function ), which---up to a phase---is the square root of the well-known Husimi ``$Q$" representation. We express the $K$ function of any $N$-mode Gaussian state as a function of its covariance matrix and displacement vector, and also that of a general continuous-variable cluster state in terms of the modal squeezing and graph topology of the cluster. This formalism lets us characterize the non Gaussian state left over when one measures a subset of modes of a Gaussian state using photon number resolving detection, the fidelity of the obtained non-Gaussian state with any target state, and the associated heralding probability, all analytically. We show that this probability can be expressed as a Hafnian, re-interpreting the output state of a circuit claimed to demonstrate quantum supremacy termed Gaussian boson sampling. As an example-application of our formalism, we propose a method to prepare a two-mode coherent-cat-basis Bell state with fidelity close to unity and success probability that is fundamentally higher than that of a well-known scheme that splits an approximate single-mode cat state---obtained by photon number subtraction on a squeezed vacuum mode---on a balanced beam splitter. This formalism could enable exploration of efficient generation of cat-basis entangled states, which are known to be useful for quantum error correction against photon loss.
\end{abstract}
\maketitle
  
\section{Introduction}
Gaussian states of bosonic modes---quantum states of light that can be prepared using quadrature squeezed light and passive linear optics---form an important set of quantum states whose elegant mathematical description~\cite{Ferraro2005} and feasibility of experimental production~\cite{Furusawa2016} make Gaussian quantum information processing a major success~\cite{Weedbrook2012}. However, it is well known that Gaussian states and Gaussian measurements (homodyne and heterodyne detection) do {\em not} constitute a universal set, i.e., resources that would allow universal quantum computation~\cite{Braun1999}. Moreover, various important protocols for quantum enhanced information processing cannot be performed when restricted to Gaussian states, Gaussian unitaries, and Gaussian measurements alone. Such no-go theorems have appeared for universal quantum computing~\cite{Bar02}, entanglement distillation~\cite{Eis02,Fiu02,Gie02}, optimal cloning of coherent states~\cite{Cer05}, optimal discrimination of coherent states~\cite{Tak08,Tsu11,Wit10}, receivers for optical communications~\cite{Namiki2014-ho}, quantum error correction~\cite{Nis09}, quantum-enhanced sensing~\cite{Gagatsos2016}, and quantum repeaters~\cite{Nam14}.

Therefore, having access to non-Gaussian states becomes imperative in pretty much any application of quantum enhanced photonic information processing. Introducing non-Gaussianity into an optical system can be challenging. For example, large $\chi^{(3)}$ non-linearities are very difficult to be implemented at optical frequencies, and obtaining a strong-enough non-Gaussian interaction through a $\chi^{(2)}$ medium with a depleted pump~\cite{Carmichael1988} is hard. An alternative way to inject non-Gaussianity is to utilize detection-induced, often probabilistic, methods such as photon number subtraction. Theoretical, numerical, and experimental studies~\cite{Dakna1997,Treps2016,Takahashi2008,Brouri2009,Marek2008,Ra2017,Dufour2017,Barnett2018,Glancy2008,Ra2019} have shown that photon subtraction on a single-mode Gaussian (squeezed vacuum) state yields approximations of coherent cat-states and have validated the non-Gaussian character of photon-subtracted multi-mode states. Further, photon subtraction has been shown to enhance entanglement~\cite{Opatrny2000,Kitagawa2006,Carlos2012}, and the fidelity of continuous-variable teleportation as was originally shown \cite{Opatrny2000} and also later studied ~\cite{Ses15}. 

Evaluating the state obtained after subtracting $m$ photons from a state $|\psi\rangle$, i.e., $\hat{a}^{m} |\psi\rangle$ using the photon number (Fock) basis $\{|n\rangle\}$, and even methods using the Husimi $Q$ representation of the state $|\psi\rangle$ lead to onerous calculations. This is because one has to calculate expressions such as $\hat{a}^{m} |n\rangle$ and $\hat{a}^{\dagger m} |\alpha\rangle$ within difficult-to-handle summations and integrals, where $\hat{a}$ is the modal photon annihilation operator. Similar difficulties apply when using the Wigner representation. Furthermore, taking into account the deviations of the photon-subtracted state from $\hat{a}^{m} |\psi\rangle$ pursuant to actual experimental methods of implementing such operation using a beamsplitter and photon number resolving (PNR) detectors, creates additional complexities. Despite photon number subtraction being a very promising tool for non-Gaussian state engineering, this analytical difficulty has come in the way of theoretical progress in the field.

In this work, we remedy the above situation by expressing the state on the coherent basis \cite{Bargmann1961,KlauderSudarshan}. Specifically, we utilize the positive $P_+$ representation of a quantum state, which is essentially expressing a general density operator in the coherent-state over-complete basis~\cite{Drummond1980}. This representation always exists unlike the Glauber-Sudarshan $P_{\textrm{GS}}$ function, which is not always defined, especially for squeezed states which are of interest in creation of cat states. The $P_+$ representation has been utilized for the numerical and analytical study of Fokker-Planck equations of dynamical systems \cite{Drummond1980,Carmichael1988,Zhu1989,Schack1991,Drummond1997,Fan1998,Olsen2009}, Ising systems \cite{Barry2008}, and single-mode quantum information analyses~\cite{Dodonov1994}. Formulas for $P_+$ have been given for Gaussian states \cite{HelstromChaptV} but only for the cases where the Glauber-Sudarshan $P_{\textrm{GS}}$ function is well defined.

We first define the $P_+$ function of an $N$-mode pure Gaussian state, which we call the $K$ function. It is a unique representation of any pure state, and can be interpreted as a square root of the Husimi $Q$ function up to a phase. The latter mathemnatical obesrvation, allow us to derive clean, closed form, and easy to use formulas for the $P_+$ representaion (called $K$ in this work), for any Gaussian state.  We begin with developing a closed-form expression of the $K$ function of a general $N$-mode Gaussian state. This lets us analytically characterize non-Gaussian states created by photon number detection and/or photon number subtraction on a subset of modes of any $N$-mode Gaussian state in an analytic integral form. We show that this reproduces---in a rather simple set of steps---the theory behind {\em Gaussian boson sampling}, where it was argued that sampling from the photon number distribution of a random $N$-mode entangled Gaussian state is a classically-hard computational task as was proven in \cite{Lund2014} and also subsequntly studied \cite{Hamilton2017,Quesada2018}. As a new example application of our formalism, we consider the problem of engineering coherent cat basis entangled cluster states. We propose a method to prepare a two-mode cat-basis Bell state by subtracting photons from both modes of a Gaussian two-mode entangled squeezed state. We show that the fidelity versus success probability trade-off of our method is higher than that of the conventional method---that of splitting an approximate single-mode cat state, obtained by photon number subtraction on a squeezed vacuum mode, on a balanced beamsplitter. The above analysis would be extremely cumbersome (and not scalable to a larger entangled state) if done in the traditional way in the photon number basis. We expect generalization of the above, to enable exploration of efficient generation of cat-basis cluster states, which have recently emerged as a very powerful resource for quantum error correction against photon losses, with applications both to photonic quantum repeaters as well as superconducting quantum computing~\cite{Michael2016-cx,Li2017-la,Albert2018-vq}.

\section{The $K$ function of a pure Gaussian state}
We work in units of $\hbar=1$, where $N$-mode vacuum state's covariance matrix (CM) is $V_0=I/2$, with $I$ being the $N$-mode identity operator. Coherent states of $N$ modes $|\vec{\alpha}\rangle$ are not mutually orthogonal. Yet they form an over-complete basis. In other words, they resolve the identity operator, viz., 
\begin{eqnarray}
\label{eq:resolution}	I=\frac{1}{(2\pi)^N}\int d^{2N} \vec{x}_{\alpha} |\vec{\alpha}\rangle \langle \vec{\alpha}|,
\end{eqnarray}
where ${\vec{x}_{\alpha}^T=(\vec{q}_{\alpha}^T,\vec{p}_{\alpha}^T)}$, and the volume element $d^{2N} \vec{x}_{\alpha}=dq_{\alpha_1}\ldots dq_{\alpha_N} dp_{\alpha_1}\ldots dp_{\alpha_N}$. We take $\alpha_i=(q_{\alpha_i}+i p_{\alpha_i})/\sqrt{2}$. Using Eq. (\ref{eq:resolution}), we can express any $N$-mode pure state $|\Psi_0\rangle$ as
\begin{eqnarray}
\nonumber	|\Psi_0\rangle&=&\frac{1}{(2\pi)^N}  \int d^{2N} \vec{x}_\alpha\langle \vec{\alpha}| \Psi_0\rangle |\vec{\alpha} \rangle\\
&=& \int d^{2N} \vec{x}_\alpha\ K(\vec{x}_\alpha) |\vec{\alpha} \rangle, \label{eq:KfunctionInt}
\end{eqnarray}
where we call $K(\vec{x}_\alpha) = 1/(2\pi)^N \langle \vec{\alpha}| \Psi_0\rangle$ the $K$-function of the state $|\Psi_0\rangle$. When compared to the $Q$ function $Q(\vec{x}_\alpha)=1/(2\pi)^N|\langle \vec{\alpha}|\Psi_0 \rangle|^2$, the $K$-function resembles something that could be called \emph{the square root} of the $Q$ function. However, one has to be careful as $\langle \vec{\alpha}|\Psi_0 \rangle$ is a complex number and its square root will contain a phase that if omitted will produce wrong results since it depends on $\vec{x}_{\alpha}$. 

Let us assume that $|\Psi_0\rangle$ is a zero-mean Gaussian state, such that $Q(\vec{x}_\alpha)=1/(2\pi)^N\langle \vec{\alpha}|\Psi_0 \rangle \langle \Psi_0|\vec{\alpha}\rangle$ is a Gaussian function. In order to calculate the $K$ function, one must break up the Gaussian $Q$ function's exponent into two conjugate parts, yielding a Gaussian $K$ function. This step becomes easier if instead of working with Cartesian coordinates $(\vec{q}_{\alpha},\vec{p}_{\alpha})$ we move to complex coordinates $(\vec{\alpha},\vec{\alpha}^*)$ with a $\pi/4$ phase space rotation. After we finish the calculation we rotate back to Cartesian coordinates. 

Let us now consider a general $N$-mode Gaussian pure state $|\Psi\rangle = D(\vec{\beta})|\Psi_0\rangle$, where $D(\vec{\beta})$ is the displacement operator. With $|\Psi_0\rangle$ expressed in its $K$-function form~\eqref{eq:KfunctionInt}, it is straightforward to evaluate the $K$-function of $|\Psi\rangle$ since $D(\vec{\beta}) \ket{\vec \alpha} = \exp\left(\vec{\beta} \vec{\alpha^*}-\vec{\beta^*} \vec{\alpha}\right) |{{\vec \alpha} + {\vec \beta}}\rangle$. 

Using the above method, we show that any $N$-mode pure Gaussian state with CM $V$ and displacement vector $\vec{x}^T_{\beta}=(\vec{q}^T_{\beta},\vec{p}^T_{\beta})$~\footnote{we work in the $qqpp$ representation, i.e., the upper left (lower right) block of the CM concerns position (momentum), while the off-diagonal blocks hold information of correlations thereof.} can be written as follows (see App. Sec. \ref{App:KnoDisp} and \ref{App:KwithDisp} for the complete derivation),
\begin{eqnarray}
\label{eq:PsiWithDisplacements}|\Psi\rangle = \int d^{2N} \vec{x}_\alpha\ K(\vec{x}_\alpha) G(\vec{x}_\alpha,\vec{x}_{\beta})  |\vec{\alpha} \rangle,
\end{eqnarray}
where,
\begin{eqnarray}
\label{eq:Kfunction} K(\vec{x}_{\alpha})&=&\frac{\exp\left[-\frac{1}{2} \vec{x}_{\alpha}^T \mathcal{B} \vec{x}_{\alpha} \right]}{(2\pi)^N(\det \Gamma)^{1/4}} ,\\
\label{eq:Gfunction} G(\vec{x}_\alpha,\vec{x}_{\beta})&=&\exp\left[\frac{1}{4} \left(\vec{x}_{\alpha}^T\ \vec{x}_{\beta}^T\right) \mathcal{D} \left(\vec{x}_{\alpha}\ \vec{x}_{\beta}\right) \right],
\end{eqnarray}
with $\Gamma=V+I/2$, and
\begin{eqnarray}
	\mathcal{B}&=&\frac{1}{2}
	\begin{pmatrix}
	A + \frac{i}{2}\left(C+C^T\right) & C - \frac{i}{2}\left(A-B\right) \\
	C^T - \frac{i}{2}\left(A-B\right) & B - \frac{i}{2}\left(C+C^T\right)
	\end{pmatrix},\\
	\mathcal{D}&=&
	\begin{pmatrix}
	0 & 2\mathcal{B}+\mathcal{X} \\
	2\mathcal{B}-\mathcal{X}& -2\mathcal{B}
	\end{pmatrix},\\
	\mathcal{X}&=&
	\begin{pmatrix}
	I & iI \\
	-iI& I
	\end{pmatrix},
\end{eqnarray}
where $A=A^T, B=B^T,\ C$ are defined as the blocks of the CM $\Gamma$ defined as follows~\footnote{Since the CM $V$ is symmetric, $\Gamma$ and $\Gamma^{-1}$ will be symmetric.}:
\begin{eqnarray}
	\Gamma^{-1}=
	\begin{pmatrix}
	A & C\\
	C^T & B
	\end{pmatrix}.
\end{eqnarray}

\section{Photon subtraction from a general multi-mode Gaussian state.}
Subtraction of $m$ photons from a single-mode quantum state $|\psi\rangle$ can be implemented by transmitting $|\psi\rangle$ through a beam splitter of transmissivity $\tau$ (chosen to be close to $1$) while detecting the low-transmissivity output of the beam splitter with a PNR detector. If the detector registers $m$ photons, the transmitted state projects to $\mathcal{P}_{-{m}}\left[|\psi\rangle\right]$, which is an approximation of the $m$-photon subtracted state $\hat{a}^{m}|\psi\rangle$. Since $\hat{a}^{m}$ is not a unitary, photon subtraction only succeeds probabilistically. 

Let us consider subtracting a vector $\vec{m} = (m_1, \ldots, m_N)$ photons from an $N$-mode pure Gaussian state $|\Psi\rangle$ using an array of beam splitters of transmissivities $\tau_i$, and PNR detectors. The post-subtraction state will be denoted $\mathcal{P}_{-\vec{m}}[|\Psi\rangle]$, implying $m_i$ photons were subtracted from the $i$-th mode, $i=1,2,\ldots,N$. Using the $K$ function of $|\Psi\rangle$~\eqref{eq:PsiWithDisplacements}, we see that $\mathcal{P}_{-\vec{m}}$ acts only on the coherent states (see App. Sec. \ref{App:Sub}), i.e., $\mathcal{P}_{-\vec{m}}[|\vec{\alpha}\rangle]$, which assumes a simple form,
$\mathcal{P}_{-\vec{m}}[|\vec{\alpha}\rangle] = \prod_{i=1}^{N} c_i |\alpha_i\sqrt{\tau_i}\rangle,\ c_i = [\left(-\sqrt{1-\tau_i}\right)^{m_i}]/[\sqrt{m_i!}]\alpha^{m_i} e^{-(1-\tau_i)|\alpha_i|^2/2}$.

The photon subtracted state $|\Psi_{-\vec{m}}\rangle$ is given as:
\begin{eqnarray}
\nonumber	|\Psi_{-\vec{m}}\rangle&=& \frac{1}{\sqrt{P}} \prod_{i=1}^N \frac{\left(-\sqrt{1-\tau_i}\right)^{m_i}}{\sqrt{{m_i}!}} \int d^{2N} \vec{x}_\alpha\ K(\vec{x}_\alpha) \\
\nonumber & \times&G(\vec{x}_\alpha,\vec{x}_\beta)e^{-\frac{(1-\tau_i)}{4}|\vec{x}_{\alpha}|^2}\\
\label{eq:PsiMinusm}& \times& \left(\frac{q_{\alpha_i}+i p_{\alpha_i}}{\sqrt{2}}\right)^{m_i} |\sqrt{\tau_i}\vec{\alpha} \rangle,
\end{eqnarray}
where $P=\langle \Psi_{-\vec{m}}|\Psi_{-\vec{m}}\rangle$ is the probability of success of the $N$-mode vector photon subtraction. $P$ is a $4N$-dimensional integral with the elementary volume $d^{2N}\vec{x}_{\alpha}d^{2N}\vec{x}_{\gamma}$ ($\vec{x}_{\gamma}$ are the coordinates of $\langle \Psi_{-\vec{m}}|$), with a Gaussian kernel, and polynomial terms $(q_{\alpha_i}+i p_{\alpha_i})^{m_i}(q_{\gamma_i}-i p_{\gamma_i})^{m_i}$. This kind of integrals can be analytically calculated (see App. Sec. \ref{App:Prob}).

If one wishes to use photon subtraction to produce a desired non-Gaussian multimode entangled state $|C\rangle$ (for example a cat-basis Bell state that we consider later), one can evaluate analytically the fidelity $F=|\langle C|\Psi_{-\vec{m}}\rangle|^2$ between the desired state $|C\rangle$ and the actual state obtained $|\Psi_{-\vec{m}}\rangle$ if $|C\rangle$ is expressed in its $K$ function form. For cat states for example, which are superpositions of coherent states $|\vec{\gamma}\rangle$, the fidelity calculation will require us to calculate the amplitude $|\langle \vec{\gamma}|\Psi_{-\vec{m}}\rangle|$, which again is a $4N$-dimensional integral, with Gaussian kernel and polynomial terms $(q_{\alpha_i}+i p_{\alpha_i})^{m_i}$, which can be analytically calculated (see App. Sec.\ref{App:Fid}).

For the rest of this paper we will restrict our attention to zero-mean states, to keep the exposition simple. Including non-zero means is a trivial extension. Further, we will assume that all the beamsplitters employed for photon subtraction on an $N$-mode Gaussian state have the same transmissivity, $\tau$. 

\section{Gaussian boson sampling and non-Gaussian state engineering}
Consider a pure $N$-mode Gaussian state $|\Psi\rangle$, the first $M < N$ modes of which are detected using PNR detectors, obtaining the outcome ${\vec n} = (n_1, \ldots, n_M)$. It is simple to show that the resulting state $|\Phi\rangle$ on the unmeasured modes is given by (see App. Sec. \ref{App:CondNorm}),
\begin{eqnarray}
\nonumber |\Phi\rangle &=& \frac{1}{\sqrt{P_M}} \prod_{i=1}^{M}\frac{1}{\sqrt{2^{n_i}n_i!}} \int d^{2N}\vec{x}_{\alpha} K(\vec{x}_{\alpha}) e^{-\frac{1}{4}x_{\alpha_i}^2}\\
\label{Phi} && \times (q_{\alpha_i}+i p_{\alpha_i})^{n_i}|\alpha_{M+1},\ldots, \alpha_{N}\rangle,
\end{eqnarray}
where we used $|\alpha\rangle = \exp(-|\alpha|^2/2)\sum_n \alpha^n/(\sqrt{n!})|n\rangle$. The probability $P_M$ of detecting the photon number pattern $\vec n$ and hence heralding the state $|\Phi\rangle$, can be calculated by setting $\langle \Phi | \Phi \rangle = 1$. 

Gaussian boson sampling is the special case of $M=N$, where all $N$ modes are detected~\cite{Hamilton2017,Quesada2018}. The success probability of detecting a photon-number pattern $\vec n$, $P_{\vec{n}}=|\langle \vec{n}|\Psi\rangle|^2=|\langle n_1\ldots n_N|\Psi\rangle|^2$ can be evaluated using our formalism, and shown to be (see App. Sec. \ref{App:Distr}),
\begin{eqnarray}
\label{eq:Pn}	P_{\vec{n}}= \frac{1}{\det \mathcal{H} \sqrt{\det\Gamma} \prod\limits_{i=1}^{N}n_i!2^{n_i}}\big|\mathcal{I}_{\vec{n}}\big|^2,
\end{eqnarray}
where,
\begin{eqnarray}
\label{IN}	\mathcal{I}_{\vec{n}} &=&\int d^{2N}\vec{x}_\alpha R(\vec{x}_{\alpha})  \prod\limits_{i=1}^{N}(q_{\alpha_i}+i p_{\alpha_i})^{n_i},\\
\label{Rdistr}	R(\vec{x}_{\alpha}) &=&  \frac{\sqrt{\det \mathcal{H}}}{(2\pi)^{N}}  e^{-\frac{1}{2}\vec{x}_\alpha^T \mathcal{H} \vec{x}_\alpha},
\end{eqnarray}
and $\mathcal{H}=\mathcal{B}+I/2$. Since $\mathcal{H}=\mathcal{H}^T$ and its real part is positive definite (see App. Sec. \ref{App:H}), Eq. (\ref{Rdistr}) is a proper Gaussian distribution. Therefore, Eq. (\ref{IN}) is the mean value $\langle f_1^{n_1}\ldots f_N^{n_N} \rangle$, where $f_i=q_{\alpha_i}+ip_{\alpha_i}$, under the distribution of Eq. (\ref{Rdistr}). Using Wick's theorem \cite{Zvonkin1997,Luque2002} we can write it as,
\begin{eqnarray}
\label{Hafnian}	\mathcal{I}_{\vec{n}} = \left\{
\begin{array}{ll}
0&\Sigma=\textrm{odd},\\
\textrm{Hf}\left(F\right)&\Sigma= \textrm{even},
\end{array}
\right.
\end{eqnarray}
where $\Sigma=\sum_{i=1}^{N}n_i$ and $\textrm{Hf}\left(F\right)$ is the Hafnian of the matrix $F$ with elements $F_{ij}=\langle f_i f_j \rangle,\ 1\leq i,j\leq \Sigma$.

\section{Photon subtraction from multi-mode squeezed cluster states}
Continuous variable (CV) quantum computing is a field that explores the use of multimode entangled squeezed states for all-photonic quantum computing. Such Gaussian cluster states of thousands of modes have been prepared experimentally~\cite{Chen2014-kf,Zhang2017-if,Yoshikawa2016-aa}. It is known however that Gaussian cluster states by themselves are not a resource sufficient for universal quantum processing. Photon number detection being the most practical ``de-Gaussification" tool, and given it is known that approximate cat states can be prepared using photon number subtraction from a single-mode squeezed vacuum, we will explore the creation of cat-basis cluster (graph) states by photon number subtraction on Gaussian cluster states.

Let us consider the Gaussian graph state $|G\rangle$ which is the result of the unitary evolution of an $N$-mode vacuum state under the unitary $\hat{U}_r=\exp{(-i r \hat{H})}$ whose generating Hamiltonian is,
\begin{eqnarray}
\label{eq:Hamiltonian1} \hat{H}=-\frac{i}{2}\sum_{i,j}^{N} G_{ij} \left(\hat{a}^\dagger_i \hat{a}^\dagger_j-\hat{a}_i \hat{a}_j\right),
\end{eqnarray}
where $\hat{a}_i$ and $\hat{a}_i^\dagger$ are the annihilation and creation operators of the $i$-th mode respectively. The state $|G\rangle$ is a squeezed entangled state among its $N$ modes. The information about which modes are entangled is described by the graph (a symmetric matrix) $G$. We assume that the squeezing parameter $r>0$ is the same for all modes~\footnote{This assumption can be dropped just by having a symmetric matrix $G$ whose elements are the different values of squeezing parameter.}. In the limit $r\to \infty$, $|G\rangle$ is a continuous variable cluster state if $G$ is a full rank matrix~ \cite{Menicucci2011}. For the same $r$, we will consider a matrix $G$ which is its own inverse, i.e., $G^2=I$. Under this assumption on $G$, we show that (see App. Sec. \ref{App:B}),
\begin{eqnarray}
\label{selfinverseG}	\mathcal{B}=\frac{1}{2}I+\frac{1}{2}\tanh r \begin{pmatrix}
	-G & i G\\
	i G & G
	\end{pmatrix}.
\end{eqnarray} 

To demonstrate the power of our method, as a first example we consider a two-mode squeezed vacuum state (TMSV), from which we subtract five photons per mode (ten in total). We calculate the photon subtracted state $|\Psi_{-5,-5}\rangle$, the probability of success $P_{5}$, and the fidelity $F_{5}=|\langle C|\Psi_{-5,-5}\rangle|^2$, where
\begin{eqnarray}
\label{CCS}	|C\rangle = \frac{1}{N_{+}}\left(|\gamma,\gamma\rangle+|-\gamma,-\gamma\rangle\right),
	\end{eqnarray}
with normalization $|N_{+}|^2 = 2[1+e^{-2(q_{\gamma}^2+p_{\gamma}^2)}]$. We compare the state $|\Psi_{-5,-5}\rangle$ with the specific state of Eq. (\ref{CCS}), because both states are parity $(-1)^{\hat{n}_1+\hat{n}_2}$ eigenstates with eigenvalue $+1$.
If the $K(\vec{x}_{\alpha})$ function is known, then the state $|\Psi_{-5,-5}\rangle$ is known from Eq. (\ref{eq:PsiMinusm}) for zero displacement. The only thing required to find the $K(\vec{x}_{\alpha})$ is the matrix $\mathcal{B}$ \footnote{The matrix $\Gamma^{-1}$ can be easily found to be $\Gamma^{-1}=\mathcal{B}+\mathcal{B}^\dagger$, from that we calculate $\det \Gamma = \cosh^4 r$.}, which is given by Eq. (\ref{selfinverseG}) for,
\begin{eqnarray}
	G=\begin{pmatrix}
	0 & 1\\
	1 & 0
	\end{pmatrix},
\end{eqnarray}
which describes the graph corresponding to the TMSV, as can also be seen by Eq. (\ref{eq:Hamiltonian1}).
The probability $P_{5}$ and the fidelity $F_{5}$ are given by:
\begin{eqnarray}
P_{5} &=& \frac{(1-\tau^2)^{10}\tanh^{10} r}{\cosh^2 r} p(\mu),\,{\text{and}}\\
F_{5}&=&\frac{2e^{-(q_{\gamma}^2+p_{\gamma}^2-\frac{z}{2})}(1-\tau)^5\tanh^2 r}{[1+e^{-2(q_{\gamma}^2+p_{\gamma}^2)}]\sqrt{P_{5}}(\det \Gamma)^{\frac{1}{4}}} w(z),
\end{eqnarray}
where $p(\mu) = [(1+\mu^2)(1+24\mu^2+76\mu^2+24\mu^6+\mu^8)]/[(1-\mu^2)^{11}]$, $w(z) = 1+(5z)/(2)+(5z^2)/(4)+(5z^3)/(24)+(5z^4)/(384)+(z^5)/(3840)$, $\mu=\tau \tanh r$, and $z=(q_{\gamma}-i p_{\gamma})\mu$. For example for $q_{\gamma}=0.5$, $p_{\gamma}=0$, $\tau=0.01$, and $r=1$ we get $P_{5}=0.025$ and $F_{5}=0.979$. Note that in the above example, the analytical complexity would not have changed if we decided to subtract more (e.g., $10$ photons) from each mode, whereas a traditional Fock basis calculation would become completely intractable.

\begin{figure}[t]
	\centering
	\includegraphics[scale=0.4]{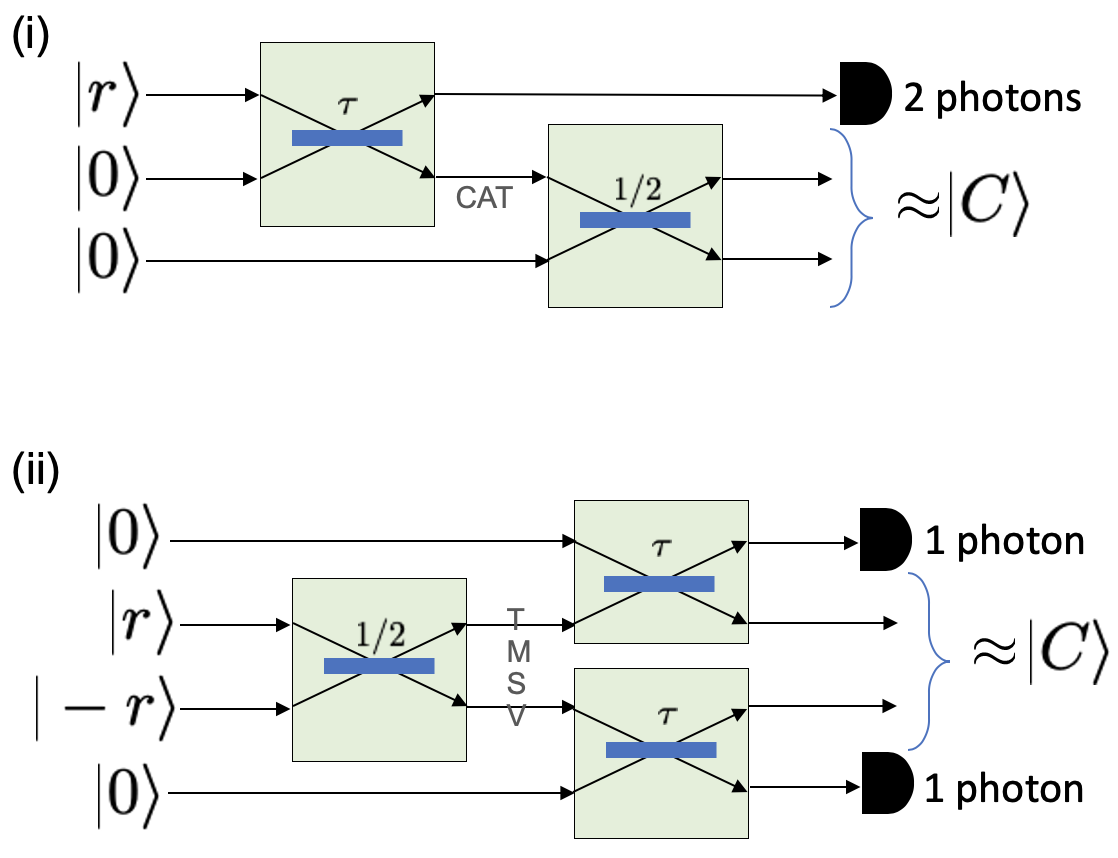}
	\caption{Scenario (i): Two-photon subtraction from a single-mode squeezed state creates a state close to a single mode coherent cat state. This when split on a balanced beam splitter, creates a state that approximates the two-mode coherent cat-basis entangled state $|C\rangle$. Scenario (ii): An approximation to $|C\rangle$ is created by subtracting one photon from each mode of a two-mode squeezed vacuum state.}
	\label{fig:scheme}
\end{figure}

\begin{figure}[t]
	\centering
	\includegraphics[width=1\linewidth]{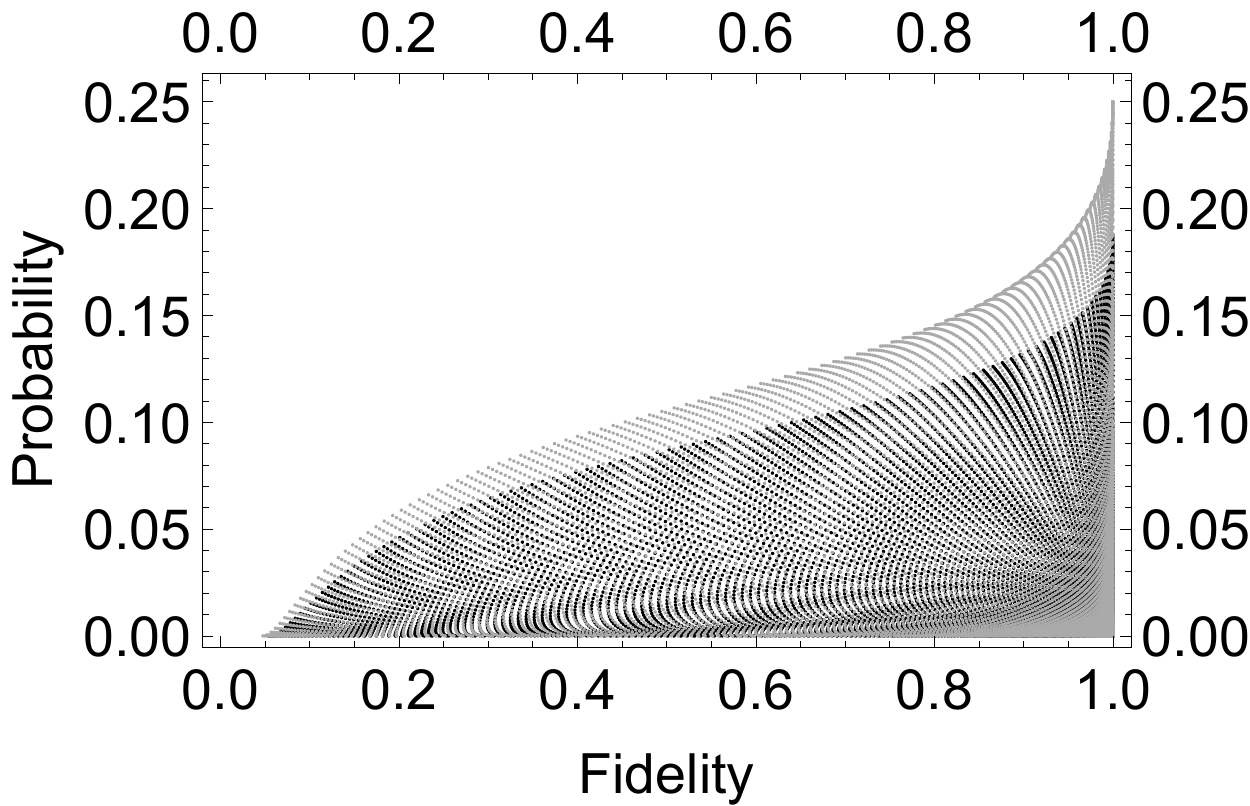}
	\caption{Black dots correspond to scenario (i) while gray dots to scenario (ii). Each dot corresponds to $q_{\gamma}=0.1,\ p_{\gamma}=0$ and $r,\ \tau$ are taking values in $[0.01,1]$ with step $0.01$. Scenario (ii) is superior to scenario (i) as it can achieve higher fidelity with higher probability of success.}\label{fig:comparisons1}
	\includegraphics[width=1\linewidth]{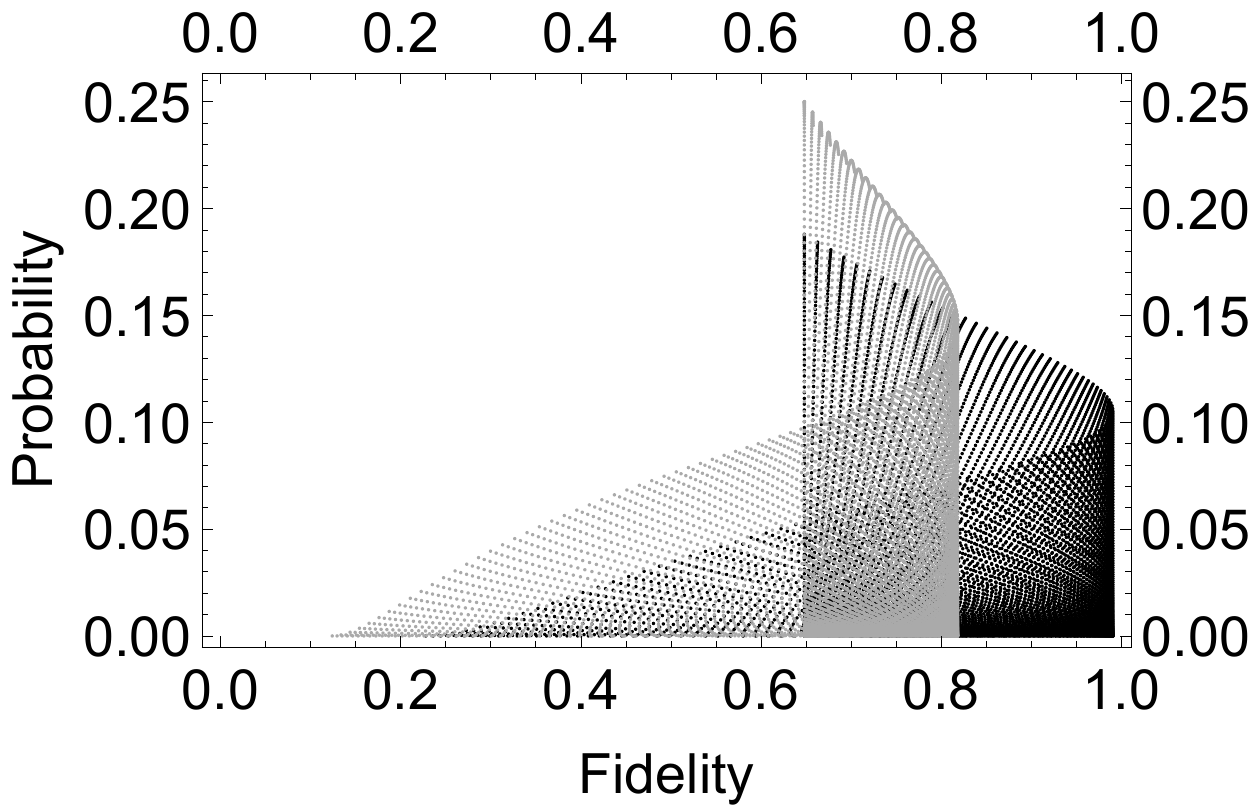}
	\caption{Black dots correspond to scenario (i) while gray dots to scenario (ii). Each dot corresponds to $q_{\gamma}=1,\ p_{\gamma}=0$ and $r,\ \tau$ are taking values in $[0.01,1]$ with step $0.01$. Scenario (i) can achieve higher fidelity. However, for high fidelity the probability of success is smaller compared to smaller coherent cat states.}
	\label{fig:comparisons2}
	\end{figure}

As a second example we consider two ways to produce the cat-basis Bell state $|C\rangle$: (i) a single-mode squeezed state from which we subtract two photons and the resulting state is known to be an approximation of the cat state $|\delta\rangle + |-\delta\rangle$, which if then split in a 50-50 beam splitter, is known to produce the state $|C\rangle$ with $\delta=\sqrt{2}\gamma$~\cite{Ralph2003}. In scenario (ii) we subtract one photon from each of the two modes of a TMSV. In both scenarios two photons are subtracted in total. Also, the beam splitter used in scenario (i) is a Hadamard gate, which if used to mix a position-squeezed state with a momentum-squeezed state we get a TMSV, see Fig. \ref{fig:scheme} We set $p_{\gamma}=0$ and we calculate the probabilities of success $P_{(i)},\ P_{(ii)}$ and the fidelities $F_{(i)},\ F_{(ii)}$ for scenarios (i) and (ii) to the desired state $|C\rangle$, as:
\begin{eqnarray}
P_{(i)} &=& \frac{(\tanh r-\mu)^2(1+2\mu^2)}{2\cosh r (1-\mu^2)^{\frac{5}{2}}},\\
P_{(ii)} &=& \frac{(\tanh r-\mu)^2(1+\mu^2)}{\cosh^2 r (1-\mu^2)^3},\\
F_{(i)} &=& \frac{2 e^{q_{\gamma}^2(1+\mu)}(q_{\gamma}^2\mu+1)^2(1-\mu^2)^{\frac{5}{2}}}{( e^{q_{\gamma}^2}+1)(1+2\mu^2)},\\
F_{(ii)} &=& \frac{e^{q_{\gamma}^2(1+\mu)}(q_{\gamma}^2\mu+2)^2(1-\mu^2)^{3}}{2( e^{2 q_{\gamma}^2}+1)(1+\mu^2)}.	
\end{eqnarray} 
Comparative results for these two scenarios are shown in Figs.~\ref{fig:comparisons1} and ~\ref{fig:comparisons2}. To produce a cat-basis Bell state $|C\rangle$ with a small amplitude, scenario (ii) is better than (i) in both fidelity and probability of success. As the amplitude of $|C\rangle$ increases, the situation begins to change: scenario (i) favors high fidelity, at the expense of smaller probability of success compared to scenario (ii). For example, for $q_{\gamma}=0.1,\ p_{\gamma}=0,\ r=0.9,\ \tau=0.4$, $P_{(i)}=0.179,\ P_{(ii)}=0.249,\ F_{(i)}=0.999,\ F_{(ii)}=0.999$ and for $q_{\gamma}=1,\ p_{\gamma}=0,\ r=0.9,\ \tau=0.4$, $P_{(i)}=0.093,\ P_{(ii)}=0.126,\ F_{(i)}=0.990,\ F_{(ii)}=0.806$.
It is of similar ease to find expressions for $P_{5},\ F_{5},\ P_{(i)},\ P_{(ii)},\ F_{(i)},\ F_{(ii)}$ for $p_{\gamma}\neq 0$ (generality is not lost by assuming real amplitude).

\section{Mixed Gaussian states}
A mixed Gaussian state $\hat{\rho}$ can be written as $\hat{\rho}=\hat{U}\hat{\rho}_{\textrm{th}}\hat{U}^\dagger$, where $\hat{\rho}_{\textrm{th}}$ is a thermal state and $\hat{U}$ is a Gaussian unitary. Using the Glauber-Sudarshan $P_{\textrm{GS}}$ function of the thermal state $P_{\textrm{GS,th}}\equiv P_{\textrm{th}}(\vec{x}_{\alpha})$, we have $\hat{\rho} = \int d^{2N} \vec{x}_{\alpha} P_{\textrm{th}}(\vec{x}_{\alpha}) |\Psi\rangle \langle\Psi|$
where $|\Psi\rangle = U |\vec{\alpha}\rangle$. The state $|\Psi\rangle$ can be expressed using Eq. (\ref{eq:PsiWithDisplacements}) and therefore $\hat{\rho}$ is expressed in the coherent-state basis as two integrals coming from $|\Psi\rangle$ and $\langle \Psi |$ are convoluted into a third integral over $\vec{x}_{\alpha}$ with $P_{\textrm{th}}(\vec{x}_{\alpha})$. 

Concerning mixed Gaussian states, things become even easier if an initial pure Gaussian sate $|\Psi_0\rangle$ goes through a pure loss channel. We remind the reader that under a pure loss channel, every mode of the state $|\Psi_0\rangle$ is coupled with $|0\rangle$ (the environment) via a beam splitter of transmittance $\tau_i$, where $i=1,\ldots,N$ counts the modes, i.e., the loss does not have to be uniform across the $N$ modes. Then the environment's output is traced out. The single-mode pure loss channel is described by the Kraus operators \cite{Ivan2011},
\begin{eqnarray}
\hat{A}_l = \sqrt{\frac{(1-\tau)^l}{l!}} \tau^{\hat{n}/2} \hat{a}^l
\label{eq:KrausPureLoss}
\end{eqnarray}
and the final state is,
\begin{eqnarray}
\hat{\rho} =\sum_{l_1,\ldots,l_n=0}^{\infty} \hat{A}_{l_1}\ldots \hat{A}_{l_N} |\Psi_0\rangle \langle\Psi_0|\hat{A}_{l_N}^\dagger\ldots \hat{A}_{l_1}^\dagger.
\label{eq:PureLoss}
\end{eqnarray}
Here we observe that if $|\Psi_0\rangle$ is expressed on the coherent basis, the operators $\tau^{\hat{n}/2} \hat{a}^l$ in Eq. (\ref{eq:KrausPureLoss}), will act on coherent states resulting to managable expressions. For further simplicity we assume the same transimittance rate $\tau$ per mode (even thouhgh this assumption can be easily dropped). The final state will be,
\begin{eqnarray}
\nonumber	\hat{\rho} &=& \int  d^{2N} \vec{\alpha}d^{2N} \vec{\beta} K(\vec{\alpha}) K^*(\vec{\beta})\times \\ \nonumber && \times  \exp\Big[-\frac{1-\tau}{2}\left(|\vec{\alpha}|^2+|\vec{\beta}|^2\right)+ \\
\label{eq:FinalStatePureLoss}&& +(1-\tau)\vec{\beta}^{*T} \vec{\alpha}\Big] |\sqrt{\tau}\vec{\alpha}\rangle \langle \sqrt{\tau} \vec{\beta}|,
\end{eqnarray}
an expresion which can be useful, for example, in an analysis of a Gaussian boson sampling with pure loss scheme.

\section{Conclusions}
We have derived a general representation of Gaussian states in the coherent-state basis, and showed that it opens the door to analytical and thorough investigations of non-Gaussian states prepared via photon subtraction and partial PNR detection of Gaussian states. We showed a simplified analysis of Gaussian boson sampling as a special case of our formalism. As a specific example application of our formalism, we considered cat-basis cluster creation by multi-mode photon subtraction on entangled Gaussian states. We showed that by subtracting photons simultaneously from both modes of a two-mode squeezed vacuum state, a coherent cat basis Bell state can be produced with higher fidelity and probability of success, compared to the well-known method of first creating a cat state via photon number subtraction of a single-mode squeezed vacuum, followed by linear-optical manipulation. The question on whether more general coherent cat basis graph states---known to be an excellent resource for quantum error correction against photon loss---can be systematically engineered from Gaussian cluster states and photon subtraction, is left open for future work. We anticipate that our formalism will prove a powerful tool for non-Gaussian cluster state engineering \cite{Arzani2018,Walschaers2018,Sabapathy2018}, which is a subject of intense interest in designing scalable solutions for all-photonic quantum computing and other forms of quantum-enhanced photonic information processing such as all-photonic quantum repeaters where photonic cluster states replace the role of quantum memories~\cite{Azuma_repeater,Pant2017-pc}, and optical-domain quantum machine learning via receivers powered with cluster states~\cite{Zhuang2019-gg}.

While preparing this paper, it came to our attention \cite{[{Private communication with Xanadu}]private} that similar phase space methods have been developed \cite{Su1902.02323,Su1902.02331} practically concurrently.
\begin{acknowledgements}
	CNG was supported by the Army Research Office (ARO) STIR program, contract number W911NF-18-1-0377. SG acknowledges Xanadu Quantum Technologies for supporting multiple useful discussions on this topic. The authors acknowledge Daiqin Su, Krishna Kumar Sabapathy, Hari Krovi, Raf Alexander, and Kaushik Seshadreesan for valuable discussions.
\end{acknowledgements}

\bibliography{PhotonSubtractionBib}

\newpage

\onecolumngrid
\appendix
\section*{Appendix}
\renewcommand{\thesubsection}{\arabic{subsection}}
\def\theequation{A\arabic{equation}}
\setcounter{equation}{0}
\renewcommand{\thefigure}{A\arabic{figure}}    
\setcounter{figure}{0}

\noindent
\subsection{Photon subtraction from a coherent state using a beam splitter}\label{App:Sub}
\begin{figure}[h]
	\centering
	\includegraphics[width=0.4\linewidth]{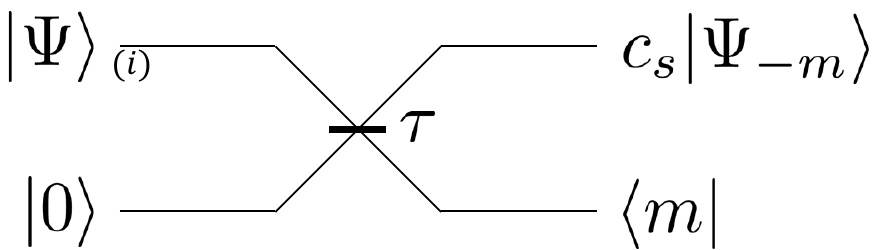}
	\caption{The $i$-th mode of a state $|\Psi\rangle$ is mixed with vacuum in a beam splitter with transmittance $\tau$. If a photon number resolution measurement registers $m$ photons in the lower output port, then $m$ photons have been subtracted from the $i$-th mode of the input state.}
	\label{fig:bs}
\end{figure}
Subtraction of $m$ photons from a mode of a state $|\Psi\rangle$ can be implemented with a beam splitter of transmittance $\tau$. The beam splitter couples the mode that the photon subtraction will take place with vacuum. Then, if the photon number resolution measurement (PNRM) registers $m$ photons, the resulting state is $|\Psi_{-m}\rangle$ as shown in Fig.~\ref{fig:bs}. Since a measurement is involved, this procedure is probabilistic and heralded. Because of the probabilistic nature of photon subtraction the final state needs to be normalized. The absolute square of the normalization is the probability of finding $m$ photons in the PNRM. This probability is also called the \emph{probability of success}.

Since we expand $|\Psi\rangle$ on coherent basis, when subtracting photons from some mode of $|\Psi\rangle$, the beam splitter will couple a coherent state with vacuum. If $\hat{a}_1,\ \hat{a}_2$ and $\hat{b}_1,\ \hat{b}_2$ are the input and output annihilation operators respectively we have,
\begin{eqnarray}
\begin{pmatrix}
\hat{b}_1 \\
\hat{b}_2
\end{pmatrix}=
\begin{pmatrix}
\sqrt{\tau} & \sqrt{1-\tau} \\
-\sqrt{1-\tau} & \sqrt{\tau}
\end{pmatrix}
\begin{pmatrix}
\hat{a}_1 \\
\hat{a}_2
\end{pmatrix}.
\end{eqnarray}
Therefore, if the global, two-mode input state is $|\alpha,0\rangle$ the final state is $|\sqrt{\tau}\alpha,-\sqrt{1-\tau}\alpha\rangle$. The conditional state on the upper output port, upon finding $m$ photons in the PNRM, is
\begin{eqnarray}
\langle m|-\sqrt{1-\tau}\alpha\rangle |\alpha\sqrt{\tau}\rangle=\frac{\left(-\sqrt{1-\tau}\right)^m}{\sqrt{m!}}\alpha^{m} e^{-(1-\tau)\frac{|\alpha|^2}{2}} |\alpha\sqrt{\tau}\rangle,
\end{eqnarray} 
therefore we can write,
\begin{eqnarray}
\hat{P}_{-m} [|\alpha\rangle]&=& c_s|\alpha_{-m}\rangle,
\end{eqnarray}
where
\begin{eqnarray}
c_s&=&\frac{\left(-\sqrt{1-\tau}\right)^m}{\sqrt{m!}}\alpha^{m} e^{-(1-\tau)\frac{|\alpha|^2}{2}},\\
|\alpha_{-m}\rangle&=&|\alpha\sqrt{\tau}\rangle
\end{eqnarray}
and the probability of success is given by $P=|c_s|^2$. Subtracting photons from a coherent state yields the same amplitude-damped coherent state $|\alpha\sqrt{\tau}\rangle$ regardless of the PNRM result. Therefore, for applications there is not much meaning in subtracting photons from coherent states. However, it is highly convenient for mathematical manipulation of photon subtraction written on coherent basis.  
\subsection{Coherent basis representation of pure Gaussian states without displacement}\label{App:KnoDisp}
We define $\vec{x}_\alpha=\begin{pmatrix}
\vec{q}_\alpha\\ \vec{p}_\alpha
\end{pmatrix}$ and we work with $\hbar=1$. Using the unit resolution on coherent states,
\begin{eqnarray}
\frac{1}{\pi^N} \int d^{2N} \vec{\alpha}\ |\vec{\alpha}\rangle \langle \vec{\alpha} | = \frac{1}{(2\pi)^N} \int d^{N} \vec{q}_\alpha d^N\vec{p}_\alpha\ |\vec{\alpha} \rangle \langle \vec{\alpha} |=\frac{1}{(2\pi)^N} \int d^{2N}\vec{x}_\alpha\ |\vec{\alpha} \rangle \langle \vec{\alpha} | =I
\end{eqnarray}
for any state $|\Psi\rangle$ we can write,
\begin{eqnarray}
\label{expansion}	|\Psi\rangle=\frac{1}{(2\pi)^N}  \int d^{2N}\vec{x}_\alpha\ \langle \vec{\alpha}| \Psi\rangle |\vec{\alpha} \rangle
= \int d^{2N}\vec{x}_\alpha\ K(\vec{x}_\alpha) |\vec{\alpha} \rangle,
\end{eqnarray}
where we define,
\begin{eqnarray}
K(\vec{x}_\alpha) = \frac{1}{(2\pi)^N} \langle \vec{\alpha}| \Psi\rangle.
\end{eqnarray}
which up to some constant is the \emph{the square root} of the $Q(\vec{x}_\alpha)$ representation,
\begin{eqnarray}
Q(\vec{x}_\alpha)=\frac{1}{(2\pi)^N}\langle \vec{\alpha}|\Psi \rangle \langle \Psi|\vec{\alpha}\rangle,
\end{eqnarray}
therefore we can write,
\begin{eqnarray}
\label{Kdef}	\frac{1}{(2\pi)^N} Q(\vec{x}_\alpha)=|K(\vec{x}_\alpha)|^2
\Rightarrow K(\vec{x}_\alpha) = \frac{1}{(2\pi)^{N/2}} Q_{1/2}(\vec{x}_\alpha),
\end{eqnarray}
such that,
\begin{eqnarray}
\label{Qhalf}	Q_{1/2}(\vec{x}_\alpha)Q_{1/2}^*(\vec{x}_\alpha)=Q(\vec{x}_\alpha).
\end{eqnarray}
Equations (\ref{Kdef}) and (\ref{Qhalf}) imply that to find $Q_{1/2}(\vec{x}_\alpha)$, we have to separate the $Q(\vec{x}_\alpha)$ representation into a product of two conjugate parts. In that way, if the state $|\Psi\rangle=|\Psi_0\rangle$ is Gaussian state with zero displacement, we can express $K(\vec{x}_\alpha)$ as a function of the states' covariance matrix (CM). The $Q(\vec{x}_\alpha)$ representation of a Gaussian state with CM $V$ is,
\begin{eqnarray}
\label{Qfunction}	Q(\vec{x}_\alpha) = \frac{1}{(2\pi)^N \sqrt{\det{\Gamma}}} \exp\left[-\frac{1}{2} \vec{x}_\alpha^T \Gamma^{-1} \vec{x}_\alpha \right],
\end{eqnarray}
where,
\begin{eqnarray}
\label{Gamma}	\Gamma = V+\frac{1}{2}I,
\end{eqnarray}
where $I$ is the identity matrix of appropriate dimensions. Any CM is a real, symmetric matrix, and as per Eq. (\ref{Gamma})  $\Gamma$ is a real, symmetric matrix. The inverse of a real, symmetric matrix is again real and symmetric, therefore in block form the matrix $\Gamma^{-1}$ is,
\begin{eqnarray}
\Gamma^{-1}=\begin{pmatrix}
A & C \\
C^T & B
\end{pmatrix},
\end{eqnarray}  
where $A=A^T$ and $B=B^T$ and $A,\ B,\ C$ real. It is more convenient if we change coordinates in the following manner,
\begin{eqnarray}
\vec{z}=R\vec{x}_{\alpha},
\end{eqnarray}
where,
\begin{eqnarray}
\vec{z}_{\alpha}&=&\begin{pmatrix}
\vec{\alpha}\\
\vec{\alpha}^*
\end{pmatrix},\\
\vec{\alpha}&=&\frac{1}{\sqrt{2}}\left(\vec{q}_{\alpha}+i \vec{p}_{\alpha}\right),\\
\vec{\alpha}^*&=&\frac{1}{\sqrt{2}}\left(\vec{q}_{\alpha}-i \vec{p}_{\alpha}\right),\\
\vec{x}_{\alpha}&=&\begin{pmatrix}
\vec{q}_\alpha\\ \vec{p}_\alpha
\end{pmatrix},\\
\label{Rmatrix}	R&=&\frac{1}{\sqrt{2}}
\begin{pmatrix}
I & I \\
-i I & i I
\end{pmatrix}.
\end{eqnarray}
Note that $R$ is unitary, i.e., $R R^\dagger = I$.

To break Eq. (\ref{Qfunction}) into two conjugate parts, we must express the term $\vec{x}_\alpha^T \Gamma^{-1} \vec{x}_\alpha$ which appears in its $\exp(.)$ as a summation of two conjugate terms. To this end we express $\Gamma$ in the $\vec{z}_{\alpha}$ basis,
\begin{eqnarray}
\vec{x}^T \Gamma^{-1} \vec{x} = \vec{z}^\dagger R^\dagger \Gamma^{-1} R \vec{z} = \vec{z}^\dagger \tilde{\Gamma}^{-1} \vec{z},
\end{eqnarray}
where,
\begin{eqnarray}
\label{GammaTilde}	\tilde{\Gamma}^{-1} = R^\dagger \Gamma^{-1} R
\end{eqnarray}
is the transformed $\Gamma^{-1}$ in the $\vec{z}_{\alpha}$ basis. From Eqs. (\ref{GammaTilde}) and (\ref{Rmatrix}) we get,
\begin{eqnarray}
\tilde{\Gamma}^{-1}=\frac{1}{2}\begin{pmatrix}
A+B-i(C-C^T) & A-B+i(C+C^T) \\
A-B-i(C+C^T) & A+B+i(C-C^T)
\end{pmatrix}=
\begin{pmatrix}
\tilde{A} & \tilde{C} \\
\tilde{C}^* & \tilde{A}^*
\end{pmatrix}.
\end{eqnarray}
Therefore we can write,
\begin{eqnarray}
\vec{z}^\dagger_{\alpha} \tilde{\Gamma}^{-1} \vec{z}_{\alpha}&=&
\begin{pmatrix} \vec{\alpha}^{* T} & \vec{\alpha}^T \end{pmatrix}
\begin{pmatrix}
\tilde{A} & \tilde{C} \\
\tilde{C}^* & \tilde{A}^*
\end{pmatrix}
\begin{pmatrix} \vec{\alpha} \\ \vec{\alpha}^*\end{pmatrix}=\\
\label{Breaking}&=&	\vec{\alpha}^{* T}\tilde{A}\vec{\alpha} + \vec{\alpha}^{* T}\tilde{C}\vec{a}^*+\vec{\alpha}^{T}\tilde{C}^*\vec{a}+\vec{\alpha}^{T} \tilde{A}^*\vec{\alpha}^* = \vec{z}^\dagger \tilde{\mathcal{B}}\vec{z}+\vec{z}^\dagger \tilde{\mathcal{B}}^\dagger\vec{z}.
\end{eqnarray}
Equation (\ref{Breaking}) shows that we can readily derive the two conjugate terms where,
\begin{eqnarray}
\label{Btilde}	\tilde{\mathcal{B}}=\frac{1}{2}
\begin{pmatrix}
\tilde{A} & \tilde{C}\\
0 & \tilde{A}^*
\end{pmatrix}.
\end{eqnarray}
Going back to Cartesian coordinates $\vec{x}_{\alpha}$ we get the matrix $\mathcal{B}$,
\begin{eqnarray}
\Gamma^{-1} = R \tilde{\Gamma}^{-1} R^\dagger =
R\tilde{\mathcal{B}}R^\dagger + R\tilde{\mathcal{B}}^\dagger R^\dagger = \mathcal{B} + \mathcal{B}^\dagger,
\end{eqnarray}
where,
\begin{eqnarray}
\label{Bmatrix}	\mathcal{B}=R\tilde{\mathcal{B}}R^\dagger=\frac{1}{2}
\begin{pmatrix}
A + \frac{i}{2}\left(C+C^T\right) & C - \frac{i}{2}\left(A-B\right) \\
C^T - \frac{i}{2}\left(A-B\right) & B - \frac{i}{2}\left(C+C^T\right)
\end{pmatrix}
\end{eqnarray}
where we have used Eqs. (\ref{Rmatrix}) and (\ref{Btilde}). Therefore given the CM $V$ of a pure Gaussian state $|\Psi_0\rangle$, we can find $\Gamma^{-1}$ and from that we can immediately write  $\mathcal{B}$ and the expansion on coherent basis is,
\begin{eqnarray}
\label{KfinalNoDis} 	K(\vec{x}_{\alpha})=\frac{1}{(2\pi)^N}\frac{1}{(\det \Gamma)^{1/4}} \exp\left[-\frac{1}{2} \vec{x}^T_{\alpha} \mathcal{B} \vec{x}_{\alpha} \right].
\end{eqnarray}
\subsection{Coherent basis representation of pure Gaussian states with displacement}\label{App:KwithDisp}
A displaced pure Gaussian state $|\Psi\rangle$ can be derived by applying a displacement $D(\vec{x}_{\beta})$ and multiple-mode squeezing $S(\vec{r})$ (phases can be absorbed into the squeezing operator) \cite{Weedbrook2012} onto a multiple-mode vacuum state $|\vec{0}\rangle$,
\begin{eqnarray}
|\Psi\rangle = D(\vec{\beta})S(\vec{r})|\vec{0}\rangle\Rightarrow |\Psi\rangle=D(\vec{\beta})|\Psi_0\rangle,
\end{eqnarray}
where $|\Psi_0\rangle$, is the state for which we worked out its coherent basis expansion in Sec.~\ref{App:KnoDisp}. Therefore we have,
\begin{eqnarray}
|\Psi\rangle&=&\frac{1}{(2\pi)^N}  \int d^{2N} \vec{x}_\alpha\ \langle \vec{\alpha}| D(\vec{\beta}) |\Psi_0\rangle |\vec{\alpha} \rangle =\frac{1}{(2\pi)^N}  \int d^{2N} \vec{x}_\alpha\ \langle \vec{0}| D(-\vec{\alpha})D(\vec{\beta}) |\Psi_0\rangle |\vec{\alpha} \rangle=\\ &=& \frac{1}{(2\pi)^N}  \int d^{2N} \vec{x}_\alpha\ \langle \vec{\alpha}-\vec{\beta}|\Psi_0\rangle \label{Psi1} e^{\frac{1}{2}\vec{\beta}\vec{\alpha}^*-\frac{1}{2}\vec{\beta}^*\vec{\alpha}}|\vec{\alpha} \rangle,
\end{eqnarray}
where in the last step we have used $D(-\vec{\alpha})D(\vec{\beta})=e^{\frac{1}{2}\vec{\beta}\vec{\alpha}^*-\frac{1}{2}\vec{\beta}^*\vec{\alpha}}D(\vec{\beta}-\vec{\alpha})$, which acts on $\langle \vec{0} |$ and therefore the sign of the displacement should be inversed. In Eq. (\ref{Psi1}), $\langle \vec{\alpha}-\vec{\beta}|\Psi_0\rangle$ is known from Eq. (\ref{KfinalNoDis}). Additionally, by defining,
\begin{eqnarray}
\label{Chi}	\mathcal{X}=
\begin{pmatrix}
I & iI \\
-iI & I
\end{pmatrix},
\end{eqnarray}
Eq. (\ref{Psi1}) is written,
\begin{eqnarray}
|\Psi\rangle&=&\frac{1}{(2\pi)^N}\frac{1}{(\det \Gamma)^{1/4}}  \int d^{2N} \vec{x}_\alpha\exp\left[-\frac{1}{2}\left(\vec{x}_{\alpha}-\vec{x}_{\beta}\right)^T\mathcal{B}\left(\vec{x}_{\alpha}-\vec{x}_{\beta}\right)\right] \exp\left(\frac{1}{4}\vec{x}_{\alpha}^T\mathcal{X}\vec{x}_{\beta}-\frac{1}{4}\vec{x}_{\beta}^T\mathcal{X}\vec{x}_{\alpha}\right)=\\
\label{Psi2}	&=&\frac{1}{(2\pi)^N}\frac{1}{(\det \Gamma)^{1/4}}  \int d^{2N} \vec{x}_\alpha \exp\left(-\frac{1}{2}\vec{x}_{\alpha}^T\mathcal{B}\vec{x}_{\alpha}\right)\exp\left[\frac{1}{4}
\begin{pmatrix}
\vec{x}_{\alpha}^T & \vec{x}_{\beta}^T 
\end{pmatrix}
\begin{pmatrix}
0 & 2\mathcal{B}+\mathcal{X} \\
2\mathcal{B}-\mathcal{X}& -2\mathcal{B}
\end{pmatrix}
\begin{pmatrix}
\vec{x}_{\alpha} \\ \vec{x}_{\beta}
\end{pmatrix}\right].
\end{eqnarray}
From Eqs. (\ref{KfinalNoDis}) and (\ref{Psi2}) we have,
\begin{eqnarray}
|\Psi\rangle = \int d^{2N} \vec{x}_\alpha\ K(\vec{x}_\alpha) G(\vec{x}_\alpha,\vec{x}_{\beta})  |\vec{\alpha} \rangle,
\end{eqnarray}
where,
\begin{eqnarray}
G(\vec{x}_\alpha,\vec{x}_{\beta})&=&\exp\left[\frac{1}{4} \left(\vec{x}_{\alpha}^T\ \vec{x}_{\beta}^T\right) \mathcal{D} \left(\vec{x}_{\alpha}\ \vec{x}_{\beta}\right) \right],
\end{eqnarray}
with
\begin{eqnarray}
\mathcal{D}&=&
\begin{pmatrix}
0 & 2\mathcal{B}+\mathcal{X} \\
2\mathcal{B}-\mathcal{X}& -2\mathcal{B}
\end{pmatrix}.
\end{eqnarray}

\subsection{Probability of success}\label{App:Prob}
The photon subtracted state is,
\begin{eqnarray}
\label{Psi3}|\Psi_{-\vec{m}}\rangle=\frac{1}{\sqrt{P}} \prod_{i=1}^N \frac{\left(-\sqrt{1-\tau_i}\right)^{m_i}}{\sqrt{{m_i}!}} \int d^{2N} \vec{x}_\alpha\ K(\vec{x}_\alpha) G(\vec{x}_\alpha,\vec{x}_\beta)e^{-\frac{(1-\tau_i)}{4}|\vec{x}_{\alpha}|^2}\left(\frac{q_{\alpha_i}+i p_{\alpha_i}}{\sqrt{2}}\right)^{m_i} |\sqrt{\tau_i}\vec{\alpha} \rangle,
\end{eqnarray}
therefore the probability of success is given by the condition $\langle\Psi_{-\vec{m}}|\Psi_{-\vec{m}}\rangle=1$. Therefore we have,
\begin{eqnarray}
\nonumber	P&=&\prod_{i=1}^N \frac{\left(1-\tau_i\right)^{m_i}}{m_i!} \int d^{2N} \vec{x}_\alpha d^{2N} \vec{x}_\gamma\ K(\vec{x}_\alpha) G(\vec{x}_\alpha,\vec{x}_\beta)K^*(\vec{x}_\gamma) G^*(\vec{x}_\gamma,\vec{x}_\beta)e^{-\frac{(1-\tau_i)}{4}\left(|\vec{x}_{\alpha}|^2+|\vec{x}_{\gamma}|^2\right)}\times\\
\label{PS}	&& \times\left(\frac{q_{\alpha_i}+i p_{\alpha_i}}{\sqrt{2}}\right)^{m_i} \left(\frac{q_{\gamma_i}-i p_{\gamma_i}}{\sqrt{2}}\right)^{m_i}\langle\sqrt{\tau_i}\vec{\gamma}|\sqrt{\tau_i}\vec{\alpha} \rangle.
\end{eqnarray}
By writing,
\begin{eqnarray}
\label{amplitude}	\langle\sqrt{\tau_i}\vec{\gamma}|\sqrt{\tau_i}\vec{\alpha} \rangle = \exp\left(-\frac{1}{4}\tau_i\vec{x}_{\alpha}^T\vec{x}_{\alpha}-\frac{1}{4}\tau_i\vec{x}_{\gamma}^T\vec{x}_{\gamma}+\frac{1}{2}\tau_i\vec{x}_{\gamma}^T\mathcal{X}\vec{x}_{\alpha}\right),
\end{eqnarray}
Eq. (\ref{PS}) gives,
\begin{eqnarray}
\nonumber	P&=&\prod_{i=1}^N \frac{\left(1-\tau_i\right)^{m_i}}{2^{m_i}m_i!} \int d^{2N} \vec{x}_\alpha d^{2N} \vec{x}_\gamma\ K(\vec{x}_\alpha) G(\vec{x}_\alpha,\vec{x}_\beta)K^*(\vec{x}_\gamma) G^*(\vec{x}_\gamma,\vec{x}_\beta)\times\\
\label{PS2}&& \times 	e^{-\frac{(1-\tau_i)}{4}\left(|\vec{x}_{\alpha}|^2+|\vec{x}_{\gamma}|^2\right)-\frac{1}{4}\tau_i\vec{x}_{\alpha}^T\vec{x}_{\alpha}-\frac{1}{4}\tau_i\vec{x}_{\gamma}^T\vec{x}_{\gamma}+\frac{1}{2}\tau_i\vec{x}_{\gamma}^T\mathcal{X}\vec{x}_{\alpha}}
 \left(q_{\alpha_i}+i p_{\alpha_i}\right)^{m_i} \left(q_{\gamma_i}-i p_{\gamma_i}\right)^{m_i}.
\end{eqnarray}
Equation (\ref{PS2}) is a Gaussian integral (represented by the $e^{-\frac{(1-\tau_i)}{4}\left(|\vec{x}_{\alpha}|^2+|\vec{x}_{\gamma}|^2\right)-\frac{1}{4}\tau_i\vec{x}_{\alpha}^T\vec{x}_{\alpha}-\frac{1}{4}\tau_i\vec{x}_{\gamma}^T\vec{x}_{\gamma}+\frac{1}{2}\tau_i\vec{x}_{\gamma}^T\mathcal{X}\vec{x}_{\alpha}}$, $K(\vec{x}_\alpha),\ K^*(\vec{x}_\gamma)$,  kernels) with linear terms (represented by $G(\vec{x}_\alpha,\vec{x}_\beta),\ G^*(\vec{x}_\gamma,\vec{x}_\beta)$), and polynomial terms $\left(q_{\alpha_i}+i p_{\alpha_i}\right)^{m_i} \left(q_{\gamma_i}-i p_{\gamma_i}\right)^{m_i}$. The way to calculate this analytically and efficiently, is to use the identity,
\begin{eqnarray}
\label{identity}	\left(q_{\alpha_i}+i p_{\alpha_i}\right)^{m_i} \left(q_{\gamma_i}-i p_{\gamma_i}\right)^{m_i} = \frac{d^{m_i}}{d\lambda_i^{m_i}} \frac{d^{m_i}}{d\mu_i^{m_i}}e^{\lambda_i\left(q_{\alpha_i}+i p_{\alpha_i}\right)+\mu_i\left(q_{\gamma_i}-i p_{\gamma_i}\right)}\Big|_{\lambda_i=\mu_i=0}.
\end{eqnarray}
Using Eq. (\ref{identity}), we cast Eq. (\ref{PS2}) into a Gaussian integral, i.e.,there is only an exponential and no polynomial terms, with extra lineal terms $\lambda_i\left(q_{\alpha_i}+i p_{\alpha_i}\right)+\mu_i\left(q_{\gamma_i}-i p_{\gamma_i}\right)$ in the exponential. Then one should take the $m_i-$order derivatives on the result of the Gaussian integral with respect to $\lambda_i$ and $\mu_i$ at $\lambda_i=\mu_i=0$.
\subsection{Fidelity}\label{App:Fid}
For any state of the form,
\begin{eqnarray}
|\phi\rangle = \sum_{i} c_i |\vec{\gamma}^{(i)}\rangle,
\end{eqnarray}
where $\sum_i |c_i|^2=1$ and $|\vec{\gamma}^{(i)}\rangle = |\gamma^{(i)}_1 \gamma^{(i)}_1\ldots \gamma^{(i)}_N\rangle$. Note that a special example of $|\phi\rangle$ is the coherent cat state (CCS) used in the main paper. The fidelity $F=|\langle \phi|\Psi_{-\vec{m}}\rangle|^2$ requires the calculation of $\langle \vec{\gamma}|\Psi_{-\vec{m}}\rangle$. From Eq. (\ref{Psi3}) we have,
\begin{eqnarray}
\nonumber \langle \vec{\gamma}|\Psi_{-\vec{m}}\rangle &=& \frac{1}{\sqrt{P}} \prod_{i=1}^N \frac{\left(-\sqrt{1-\tau_i}\right)^{m_i}}{\sqrt{{m_i}!}} \int d^{2N} \vec{x}_\alpha\ K(\vec{x}_\alpha) G(\vec{x}_\alpha,\vec{x}_\beta)e^{-\frac{(1-\tau_i)}{4}|\vec{x}_{\alpha}|^2}\times\\
\label{FidAmpl}&& \times \left(\frac{q_{\alpha_i}+i p_{\alpha_i}}{\sqrt{2}}\right)^{m_i} \langle \vec{\gamma}|\sqrt{\tau_i}\vec{\alpha} \rangle,
\end{eqnarray}
where the probability of success $P$ should be calculated first as per Sec. \ref{App:Prob}. We have,
\begin{eqnarray}
\label{amplitude2}	\langle\vec{\gamma}|\sqrt{\tau_i}\vec{\alpha} \rangle = \exp\left(-\frac{1}{4}\tau_i\vec{x}_{\alpha}^T\vec{x}_{\alpha}-\frac{1}{4}\vec{x}_{\gamma}^T\vec{x}_{\gamma}+\frac{1}{2}\sqrt{\tau_i}\vec{x}_{\gamma}^T\mathcal{X}\vec{x}_{\alpha}\right),
\end{eqnarray}
therefore Eq. (\ref{FidAmpl}) is written as,
\begin{eqnarray}
\nonumber \langle \vec{\gamma}|\Psi_{-\vec{m}}\rangle &=& \frac{1}{\sqrt{P}} \prod_{i=1}^N \frac{\left(-\sqrt{1-\tau_i}\right)^{m_i}}{\sqrt{2^{m_i}{m_i}!}} \int d^{2N} \vec{x}_\alpha\ K(\vec{x}_\alpha) G(\vec{x}_\alpha,\vec{x}_\beta)e^{-\frac{(1-\tau_i)}{4}|\vec{x}_{\alpha}|^2-\frac{1}{4}\tau_i\vec{x}_{\alpha}^T\vec{x}_{\alpha}-\frac{1}{4}\vec{x}_{\gamma}^T\vec{x}_{\gamma}+\frac{1}{2}\sqrt{\tau_i}\vec{x}_{\gamma}^T\mathcal{X}\vec{x}_{\alpha}}\times\\
\label{FidAmpl2}&& \times \left(q_{\alpha_i}+i p_{\alpha_i}\right)^{m_i}.
\end{eqnarray}
Equation (\ref{FidAmpl2}), similarly to $P$ in Sec. \ref{App:Prob}, is a Gaussian integral with linear terms, and polynomial terms $\left(q_{\alpha_i}+i p_{\alpha_i}\right)^{m_i}$ which can be injected into the exponential of Eq. (\ref{FidAmpl2}) by using the identity,
\begin{eqnarray}
\label{identity2}	\left(q_{\alpha_i}+i p_{\alpha_i}\right)^{m_i}= \frac{d^{m_i}}{d\lambda_i^{m_i}} e^{\lambda_i\left(q_{\alpha_i}+i p_{\alpha_i}\right)}\Big|_{\lambda_i=0}.
\end{eqnarray}
That way Eq. (\ref{FidAmpl2}) will become a Gaussian integral, upon which we take $m_i-$order derivatives with respect to $\lambda_i$ at $\lambda_i=0$. 
\subsection{The conditional state and its normalization}\label{App:CondNorm}
We set zero displacements, therefore we work with the $N-$mode Gaussian state $|\Psi_0\rangle$. Upon finding a pattern $\{n_1,\ldots,n_M\},\ M<N$ at the photon number resolution measurements (PNRM) at each one of the $M$ modes, the conditional state $|\Phi\rangle$ is,
\begin{eqnarray}
\nonumber|\Phi\rangle &=&\frac{1}{\sqrt{P_M}} \langle n_1,\ldots,n_M|\Psi_0\rangle\\
&=&\frac{1}{\sqrt{P_M}} \prod_{i=1}^{M}\frac{1}{\sqrt{2^{n_i}n_i!}}\int d^{2N}\vec{x}_{\alpha} K(\vec{x}_{\alpha})e^{-\frac{1}{4}x_{\alpha_i}^2}
\label{Phi} (q_{\alpha_i}+i p_{\alpha_i})^{n_i}|\alpha_{M+1},\ldots, \alpha_{N}\rangle.
\end{eqnarray}
The probability $P_M$ is given by the normalization $\langle\Phi|\Phi\rangle=1$,
\begin{eqnarray}
\nonumber P_M	&=& \prod_{i=1}^{M}\frac{1}{2^{n_i}n_i!}\int d^{2N}\vec{x}_{\alpha} d^{2N}\vec{x}_{\gamma} K(\vec{x}_{\alpha}) K^*(\vec{x}_{\gamma}) e^{-\frac{1}{4}|x_{\alpha}|^2-\frac{1}{4}|x_{\gamma}|^2+\sum\limits_{k,l=M+1}^{N} x_{\gamma_k}\mathcal{X}_{kl} x_{\alpha_l}}\times\\
\label{PM} && \times (q_{\alpha_i}+i p_{\alpha_i})^{n_i} (q_{\gamma_i}-i p_{\gamma_i})^{n_i}
\end{eqnarray}
where we have used $\langle \gamma | \alpha \rangle = \exp(-|\gamma|^2/2-|\alpha|^2/2+\gamma^* \alpha)$, $|\vec{x}_{\alpha,\gamma}|^2=\sum_{k=1}^{N} x^2_{\alpha_i,\gamma_i}$, and $\mathcal{X}_{kl}$ are the matrix elements of $\mathcal{X}$ of Eq. (\ref{Chi}) for dimensions $(N-M)\times(N-M)$. The same method using ancillary variables $\lambda_i$  as in Sec. \ref{App:Prob} can be applied to calculate $P_M$ of Eq. (\ref{PM}). 
\subsection{The probability distribution $P_{\vec{n}}$}\label{App:Distr}
We set zero displacements. The probability of finding a pattern $\{n_1,\ldots,n_N\}$ at each one of all the $N$ modes is,
\begin{eqnarray}
\label{Pn} P_{\vec{n}}=|\langle \vec{n}|\Psi_0\rangle|^2=|\langle n_1\ldots n_N|\Psi_0\rangle|^2	
\end{eqnarray}
From Eq. (\ref{KfinalNoDis}) and using $\langle n | \alpha \rangle = \exp(-|\alpha|^2/2) \alpha^{n*}/(\sqrt{n!})$ we get,
\begin{eqnarray}
\nonumber \langle n_1\ldots n_N|\Psi_0\rangle&=&\frac{1}{(2\pi)^N (\det \Gamma)^{1/4}}\prod_{i=1}^{M}\frac{1}{\sqrt{2^{n_i}n_i!}}\int d^{2N}\vec{x}_{\alpha} e^{-\frac{1}{2}\vec{x}_{\alpha}^T\mathcal{B}\vec{x}_{\alpha}-\frac{1}{4}\vec{x}_{\alpha}^T\vec{x}_{\alpha}}(q_{\alpha_i}+i p_{\alpha_i})^{n_i}\\
\label{Amplitude2} &=&\frac{1}{(2\pi)^N (\det \Gamma)^{1/4}}\prod_{i=1}^{M}\frac{1}{\sqrt{2^{n_i}n_i!}}\int d^{2N}\vec{x}_{\alpha} e^{-\frac{1}{2}\vec{x}_{\alpha}^T\mathcal{H}\vec{x}_{\alpha}}(q_{\alpha_i}+i p_{\alpha_i})^{n_i},
\end{eqnarray}
where $\mathcal{H}=\mathcal{B}+I/2$. As it is shown in Sec. \ref{App:H}, $\mathcal{H}$ is symmetric with positive definite real part. Therefore, the function,
\begin{eqnarray}
\label{SMRdistr}	R(\vec{x}_{\alpha}) =  \frac{\sqrt{\det \mathcal{H}}}{(2\pi)^{N}}  e^{-\frac{1}{2}\vec{x}_\alpha^T \mathcal{H} \vec{x}_\alpha}
\end{eqnarray}
represents a Gaussian distribution. In that way, Eq. (\ref{Amplitude2}) is written as,
\begin{eqnarray}
\nonumber \langle n_1\ldots n_N|\Psi_0\rangle
&=&\frac{1}{\sqrt{\det \mathcal{H}}(\det \Gamma)^{1/4}}\prod_{i=1}^{M}\frac{1}{\sqrt{2^{n_i}n_i!}}\int d^{2N}\vec{x}_{\alpha} R(\vec{x}_{\alpha})(q_{\alpha_i}+i p_{\alpha_i})^{n_i}\\
\label{Amplitude3} &=&  \frac{1}{\sqrt{\det \mathcal{H}}(\det \Gamma)^{1/4}}\frac{1}{\sqrt{2^{n_1}n_1!\ldots 2^{n_N}n_N!}}\langle f_1^{n_1} \ldots f_N^{n_N} \rangle,
\end{eqnarray}
where $f_i=q_{\alpha_i}+i p_{\alpha_i}$. Mean values of the form $\langle f_1^{n_1} \ldots f_N^{n_N} \rangle$ represent Hafnians via Wick's theorem as argued in the main paper. Equation (\ref{Amplitude3}) yields a complex number result, whose absolute squared is the probability $P_{\vec{n}}$ of Eq. (\ref{Pn}).

\subsection{$\mathcal{H}$ matrix is symmetric and its real part is positive definite}\label{App:H}
From Eq. (\ref{Bmatrix}) and given that $A^T=A$ and $B^T=B$ we can readily see that $\mathcal{B}^T=\mathcal{B}$. Therefore $\mathcal{H}=\mathcal{B}+I/2$ is symmetric as well. The real  part of $\mathcal{H}$ is,
\begin{eqnarray}
\textrm{Re}\left(\mathcal{H}\right)=\frac{1}{2}\left(\mathcal{H}+\mathcal{H}^\dagger\right)=\frac{1}{2}\left(\mathcal{H}+\mathcal{H}^*\right)=\frac{1}{2}\left[\begin{pmatrix}
A & C \\
C^T & B
\end{pmatrix}+I\right]=\frac{1}{2}\left(\Gamma^{-1}+I\right).
\end{eqnarray}
Since any CM $V$ is positive definite, denoted as $V>0$, then $\Gamma=V+I/2>0 \Rightarrow \Gamma^{-1}>0$ since the inverse of a positive definite matrix is also positive definite. 
\subsection{$\mathcal{B}$ matrix for multiple modes squeezed states}\label{App:B}
The Hamiltonian,
\begin{eqnarray}
\label{eq:Hamiltonian} \hat{H}=-\frac{i}{2}\sum_{i,j}^{N} G_{ij} \left(\hat{a}^\dagger_i \hat{a}^\dagger_j-\hat{a}_i \hat{a}_j\right)
\end{eqnarray}
generates the unitary $\hat{U}_r=\exp{(-i r \hat{H})}$ which corresponds to the symplectic matrix,
\begin{eqnarray}
S_r=
\begin{pmatrix}
e^{r G} & 0 \\
0 & e^{-r G}
\end{pmatrix}.
\end{eqnarray} 
Therefore the CM is,
\begin{eqnarray}
V=\frac{1}{2}S_r S^T_r=\frac{1}{2}
\begin{pmatrix}
e^{2 r G} & 0 \\
0 & e^{-2 r G} 
\end{pmatrix}
\end{eqnarray}
and from Eq. (\ref{Bmatrix}) we get,
\begin{eqnarray}
\label{BmatrixSq}	\mathcal{B}=\frac{1}{2}I+\frac{1}{2} \begin{pmatrix}
-\tanh G r & i \tanh G r\\
i \tanh G r & \tanh G r
\end{pmatrix}.
\end{eqnarray}
In Eq. (\ref{BmatrixSq}) the matrix $G=G^T$ is in the argument of $\tanh(.)$ which denotes,
\begin{eqnarray}
\label{tanh}	\tanh G r = \frac{e^{2rG}-I}{e^{2rG}+I}.
\end{eqnarray}   
For a self-inverse matrix $G=G^{-1}$, i.e., $G^2=I$, we expand $e^{2rG}$ in Taylor series. That way we get,
\begin{eqnarray}
\label{e1}	e^{2rG}-I &=& I \cosh 2r + G \sinh 2r-I\\
\label{e2}	e^{2rG}+I &=& I \cosh 2r + G \sinh 2r+I.
\end{eqnarray}
From Eqs. (\ref{tanh}), (\ref{e1}), and (\ref{e2}), we have,
\begin{eqnarray}
\label{tanh2}	\tanh G r = \tanh^2 r \left(I+G\frac{1}{\tanh r}\right)\left(I+G\tanh r\right)^{-1}.
\end{eqnarray}
We have that,
\begin{eqnarray}
I&=&\left(I+G\tanh r\right) \left(I-G\tanh r\right) \cosh^2 r \Rightarrow\\
\label{inverse} \Rightarrow	\left(I+G\tanh r\right)^{-1}&=&\left(I-G\tanh r\right) \cosh^2 r.
\end{eqnarray}
Equations (\ref{tanh2}) and (\ref{inverse}) give,
\begin{eqnarray}
\label{tanh3}	\tanh G r = G \tanh r.
\end{eqnarray}
From Eqs. (\ref{BmatrixSq}) and (\ref{tanh3}) we find,
\begin{eqnarray}
\mathcal{B}=\frac{1}{2}I+\frac{1}{2}\tanh r \begin{pmatrix}
-G & i G\\
i G & G
\end{pmatrix}.
\end{eqnarray}

\end{document}